\documentclass[12pt,a4paper]{article}
\usepackage[latin1]{inputenc}
\usepackage{amsmath}
\usepackage{amsmath} 
\usepackage{amsfonts}
\usepackage{amssymb}
\usepackage{float}
\usepackage{cite}
\usepackage{breqn}
\usepackage{graphics}
\usepackage{mathtools}
\usepackage{xcolor}
\usepackage{mathrsfs}
\newcommand{\plainfootnote}[1]{%
  \begingroup
  \renewcommand{\thefootnote}{}
  \footnotetext{#1}%
  \addtocounter{footnote}{-1}
  \endgroup
}
\usepackage{fontawesome}
\definecolor{lightred}{rgb}{1, 0.5, 0.5}
\usepackage{tikz}
\usepackage{adjustbox}
\DeclareGraphicsExtensions{.pdf,.png,.jpg}
\usepackage[colorlinks=true,linkcolor=blue,citecolor=red]{hyperref}
\makeatletter
\newcommand{\mathleft}{\@fleqntrue\@mathmargin0pt}
\newcommand{\mathcenter}{\@fleqnfalse}
\makeatother
\newcommand*{\SavedEqref}{}
\let\SavedEqref\eqref
\renewcommand*{\eqref}[1]{%
  \begingroup
    \hypersetup{
      linkcolor=linkequation,
      linkbordercolor=linkequation,
    }%
    \SavedEqref{#1}%
  \endgroup
}

\usepackage[a4paper, width = 170mm, top = 30mm, bottom = 25mm, 
bindingoffset = 0mm]{geometry}

\def\beq{\begin{equation}}
\def\eeq{\end{equation}}
\def\bea{\begin{eqnarray}}
\def\eea{\end{eqnarray}}

\definecolor{lightorange}{RGB}{255, 178, 102} 
\definecolor{mediumlightblue}{RGB}{100, 170, 255}   
\definecolor{mediumlightred}{RGB}{230, 80, 80}      

\begin{document}
 
\begin{center}
	{\large \bf SU(2) polarization evolution on higher-order Poincar\'e  sphere\\by using general $q$-plate
}
\vspace{0.7cm}

{\sf \small Mohammad Umar\textsuperscript{\textcolor{red}{\textnormal{*}}}, P. Senthilkumaran\textsuperscript{\textcolor{blue}{\textnormal{*}}}}

\bigskip
\plainfootnote{\textcolor{red}{\faEnvelope}\, aliphysics110@gmail.com, opz238433@opc.iitd.ac.in}

\plainfootnote{\textcolor{blue}{\faEnvelope}\textsuperscript{\textcolor{blue}{\textnormal{}}} 
psenthilk@yahoo.com, psenthil@opc.iitd.ac.in}

{\em
\textsuperscript{\textcolor{red}{\textnormal{}}}Optics and Photonics Centre\\  
Indian Institute of Technology Delhi\\ 
New Delhi 110016, INDIA\\}
\vspace{0.5em}


\vspace{1.2cm}	
\noindent {\bf Abstract}		
\end{center}

\noindent
This paper investigates the rotational dynamics on the higher-order Poincar\'e sphere with the use of $q$-plate by exploring three key aspects: the topological condition, the global-local rotation, and the SU(2) polarization evolution on the sphere. The polarized light beam corresponding to this sphere and $q$-plates shares analogous topological features, characterized by azimuthal variation. We have formulated the topological condition that establishes a connection between the $q$-plate and the higher-order Poincar\'e sphere, enabling the SU(2) polarization evolution on the same higher-order Poincar\'e sphere. Leveraging this correspondence, we have shown that a single \textit{global} SO(3) rotation on the higher-order Poincar\'e sphere is a collection of multiple \textit{local} SO(3) rotations on the standard Poincar\'e sphere. SO(3) is related to SU(2) through a two-to-one surjective homomorphism, with SU(2) serving as its double cover. Moreover, we demonstrate that a general $q$-plate, defined by a continuously tunable retardance ranging from $0$ to $2\pi$ and an offset angle ranging from $0$ to $\pi/2$, provides the complete coverage on the higher-order Poincar\'e sphere.

\medskip
\vspace{1in}
\newpage
	

\section{Introduction}
The polarization states of a plane electromagnetic wave can be bijectively mapped to the surface of the unit $\textbf{S}^{2}$ sphere, providing a geometric framework that captures both the physical and mathematical essence of polarized light. This insight, originally developed by H. Poincar\'e \cite{poincare1954theorie}, led to the creation of the Poincar\'e sphere (PS), a geometric structure that allows the visualization of the state of polarization (SOP) of light. In this context, the Stokes parameters ($S_{1}, S_{2}, S_{3}$) which satisfying the relation $S_{0}^{2}=S_{1}^{2}+S_{2}^{2}+S_{3}^{2}$ function as the Cartesian coordinates on the surface of the sphere, establishing an isomorphism between the space of polarization states and points on the sphere. Each point on the PS represents a spatially homogeneous SOP, and the transformation between such states is governed by the elements of the three-parameter SU(2) group, a Lie group. These transformations are mathematically analogous to rotations on the PS, which are described by elements of the SO(3) group. This equivalence arises due to the two-to-one homomorphism that exists between the SU(2) and SO(3) groups.\\
\indent
Homogeneous waveplates, such as the quarter-wave plate (QWP) and half-wave plate (HWP), are famous members of the SU(2) group, which provides SU(2) transformations between pairs of SOPs. Implementation of SU(2) transformations by these elements corresponds to a rotation of $\pi/2$ and $\pi$ (equal to the retardance of the waveplates), respectively, on the PS. To happen this rotation, the rotation axis is parameterized by $\mathbf{k}=(\cos2\alpha, \sin2\alpha, 0)^{T}$, where $\alpha$ is the fast axis orientation of the waveplate. This parameterization shows that the rotation axis is constrained to move only in a two dimensional plane and this is because the fast axis is also contained in a two-dimensional plane, the transverse plane of the plate. Notably, both the beam, represented by a point on the PS, and the waveplate share a common topological characteristic of homogeneity. In the beam, this homogeneity refers to the uniformity of the SOPs across the cross-section. Similarly, in the waveplate, homogeneity implies that both the retardance and the orientation of the fast axis are spatially invariant within the transverse plane of the waveplate.\\
\indent
The PS is a well-celebrated geometric model, yet it fails to capture the complexity of singular beams and similar types of light beams. These beams should be regarded as extended regions on the PS, rather than being represented by a single point.  In contrast, the higher-order Poincar\'e sphere, a novel form of the $\textbf{S}^{2}$ sphere, provides a more suitable framework for the higher-order solutions of Maxwell's equations, accommodating spatially inhomogeneous beams. In essence, HOPS generalizes the standard PS. G. Millione et al. \cite{milione2011higher} introduced this sphere, where inhomogeneous light beams of constant ellipticity are represented as points on its surface. While the basis states on the PS are the right- and left-circularly polarized states and are defined solely by spin angular momentum, the basis states on the HOPS are the right- and left-circularly polarized vortex beams of equal and opposite topological charges. Each order of the HOPS is linked to a unique polarization topology, which is quantitatively described by the Poincar\'e-Hopf (PH) index \cite{freund2002polarization}, given by 
\begin{equation}
\eta = \frac{1}{2\pi} \oint \nabla \gamma \cdot dl,
\label{eta_01}
\end{equation}
where $\gamma$ represents the azimuth/orientation of the polarization elliptically/linearly polarized light. The line integral is evaluated over a closed contour that encircles a polarization singularity \cite{ruchi2020phase, senthilkumaran2024singularities}, where $\gamma$ becomes undefined. The PH index can take a positive or negative sign, indicating the handedness of the azimuthal rotation of the SOP in the vicinity of the singularity.\\
\indent
The advent of HOPS has garnered significant attention within the optics community. Numerous studies have explored various aspects of HOPS beams, including their generation \cite{naidoo2016controlled, chen2014generation, ji2023controlled, liu2024generation, yao2022generation}, as well as their polarization evolution via metasurfaces \cite{liu2014realization} and photonic Dirac points \cite{xu2021polarization}. Research has also focused on the detection of HOPS beams \cite{yao2022generation, yao2023quantitative}, Stokes polarimetry for HOPS \cite{bansal2023stokes}, and the exploration of the spatiotemporal higher-order Poincar\'{e} sphere \cite{fickler2024higher}. Additionally, investigations into tight focusing of HOPS beams have been reported \cite{pal2024tight}. A new type of polarimetry, metasurface photonics for polarization detection clock, tailored for HOPS is also reported \cite{yang2025metasurface}. Despite the significant advancements in the study of HOPS, the theoretical description of SO(3) rotations on the HOPS remains inadequately explored. In this work, we examine the polarization evolution on the HOPS framework, specifically in the context of rotational dynamics. As rotation on the standard PS is achievable through homogeneous waveplates, a similar rotation on the HOPS can be accomplished using inhomogeneous waveplates, specifically $q$-plates. This connection between HOPS beams and $q$-plates stems from the fact that both share the same topological feature.\\
\indent
The $q$-plate \cite{marrucci2006optical, marrucci2013q, rubano2019q, delaney2017arithmetic, machavariani2008spatially, kadiri2019wavelength, bansal2020use} is also a member of the SU(2) group, an SU(2) optical birefringent element. Its topology differs from that of the QWP and HWP in the sense that, in the $q$-plate, the fast axis orientation varies azimuthally in its transverse plane. The fast axis orientation is mathematically expressed as $q\phi + \alpha_0$, where $q$ is the topological charge, which can take integer or half-integer values. The parameter $\alpha_0$ represents the offset angle of the $q$-plate, which is the fast axis orientation with respect to a given reference axis. Depending on the retardance of the $q$-plate, it behaves as a quarter-wave $q$-plate ($q^Q$-plate) or a half-wave $q$-plate ($q^{H}$-plate), corresponding to retardance values of $\pi/2$ and $\pi$, respectively.\\
\indent
The aim of this paper is to explore the rotational dynamics on HOPS. We have shown that a global SO(3) rotation on HOPS is a collection of many local SO(3) rotations on the standard PS. Additionally, we have mathematically demonstrated the topologically matched condition required to navigate on HOPS. Furthermore, this paper explores the SU(2) polarization evolution on HOPS using a \textit{general} $q$-plate. The general $q$-plate is an extension of the conventional $q$-plate, defined by a continuously tunable retardance and offset angle, ranging from $0$ to $2\pi$ and $0$ to $\pi/2$, respectively. This level of tunability is attainable through the use of modern technologies such as metamaterials, metasurfaces, and liquid crystal-based devices.\\
\indent
The paper is organized as follows: In Section \ref{section_02}, we present the mathematical framework for the HOPS beam and its corresponding representation on the HOPS. Section \ref{section_03} delves into the geometry of the $q$-plate. In Section \ref{section_04}, we derive the topologically matched condition for the $q$-plate and HOPS, in order to navigate on the same HOPS. Section \ref{section_05} demonstrates, through a geometric approach, how a single global SO(3) rotation on HOPS is comprised of multiple local SO(3) rotations on the standard PS. Furthermore, Section \ref{section_06} explores the SU(2) polarization evolution on HOPS using a generalized $q$-plate. Finally, the paper reaches its conclusion in Section \ref{section_07}.
\section{The (higher-order) Poincar\'e sphere}
\label{section_02}
Mathematically, the HOPS beams are expressed as the superposition of two beams carrying equal but opposite OAM in orthogonal spin states as 
\begin{equation}
    |\psi_{\ell}\rangle = \psi_{R}|R_{\ell}\rangle+\psi_{L}|L_{\ell}\rangle,
    \label{psi}
\end{equation}
where $|R_{\ell}\rangle$ and $|L_{\ell}\rangle$ are the orthogonal orbital angular momentum (OAM) basis states for HOPS \cite{milione2011higher, bansal2023stokes} defined as $|R_{\ell}\rangle=e^{-i\ell\phi}|R\rangle$ and $|L_{\ell}\rangle=e^{i\ell\phi}|L\rangle$.  where $|R\rangle$ and $|L\rangle$ are the orthogonal spin states defined as $|R\rangle=(\hat{\textbf{x}}-i\hat{\textbf{y}})/\sqrt{2}$ and $|L\rangle=(\hat{\textbf{x}}+i\hat{\textbf{y}})/\sqrt{2}$. The Poinca\'{e} Hopf (PH) index of the HOPS beam expressed in Eq. (\ref{eta_01}) can be written as half of the difference between the topological charge of the vortex associated with LCP and RCP, hence $\eta=\frac{1}{2}[\ell-(-\ell)]=\ell$. The term $\psi_{R}$ and $\psi_{L}$ are the complex amplitudes expressed as 
\begin{equation}
    \psi_{R} = \langle R_{\ell}|\psi_{\ell}\rangle = \cos{\left(\frac{\pi}{4}-\chi^{(\eta)}\right)}e^{-i\gamma^{(\eta)}},
\end{equation}
\begin{equation}
    \psi_{L} = \langle L_{\ell}|\psi_{\ell}\rangle = \sin{\left(\frac{\pi}{4}-\chi^{(\eta)}\right)}e^{i\gamma^{(\eta)}}.
\end{equation}

\noindent
where $\gamma^{(\eta)}$ and $\chi^{(\eta)}$ are associated with the longitude and latitude on HOPS respectively, such that $2\gamma^{(\eta)}$ and $2\chi^{(\eta)}$ correspond to the longitude and latitude coordinates on the HOPS respectively. For $\eta=0$, the Eq. (\ref{psi}) represents the homogeneously polarized light. In this case, the HOPS reduces to the standard PS, with the longitude and
\begin{figure}
\centering
\includegraphics[width=0.78\linewidth]{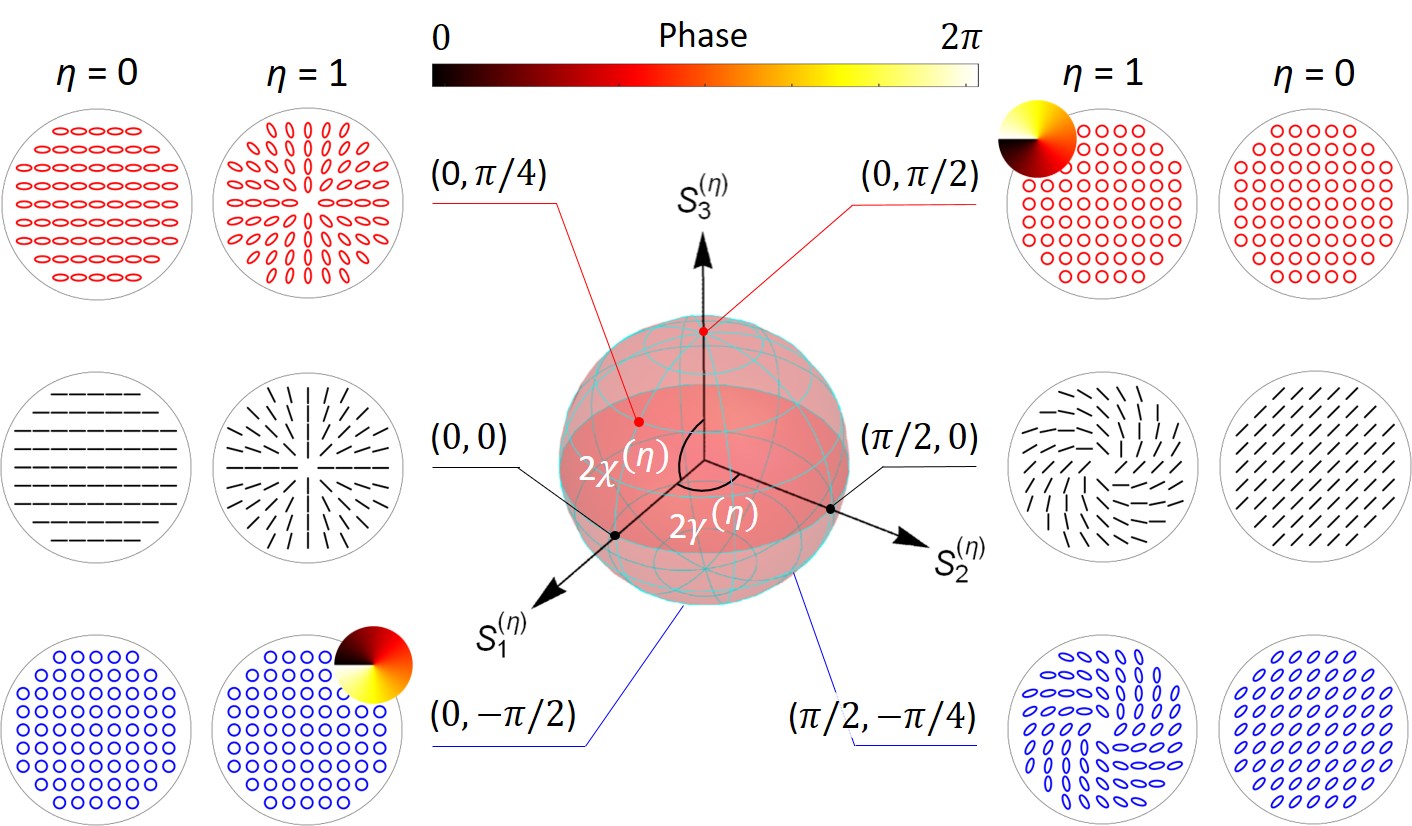}
\caption{(Color online). \textit{The geometry of the PS ($\eta=0$) and HOPS ($\eta=1$), along with their corresponding polarization distributions represented as individual points on the surface, are shown. Here, $2\gamma^{(\eta)}$ and $2\chi^{(\eta)}$ denote the longitude and latitude coordinates, respectively. Red and blue colors represent right-handed and left-handed polarization, respectively. The vortex phase corresponding to the RCP and LCP eigenstates is also shown.}}
\label{q022}
\end{figure}
latitude coordinates $2\gamma^{(0)}$ and $2\chi^{(0)}$, respectively and are related to the azimuth and ellipticity of the SOP. For $\eta\neq 0$, $2\gamma^{(\eta)}$ and $2\chi^{(\eta)}$ are related to the Pancharatnam phase and ellipticity, respectively \cite{bansal2023stokes}. On the HOPS, the equator is occupied by the cylindrical vector beam, including both radially and azimuthally polarized light, and this occurs when $|\psi_{R}| = |\psi_{L}|$ while non-equatorial points correspond to cases when $|\psi_{R}| \neq |\psi_{L}|$. The geometry of the PS and HOPS is presented in Fig. \ref{q022} for $\eta=1$.

The Stokes parameters (SPs) corresponding to the singular beam expressed in Eq. (\ref{psi}) are defined as $S_{0}^{(\eta)} = |\psi_R|^{2} + |\psi_L|^{2}$, $S_{1}^{(\eta)} = 2\texttt{Re}[\psi_{R} \psi_{L}^{*}]$, $S_{2}^{(\eta)} = 2\texttt{Im}[\psi_{R} \psi_{L}^{*}]$ and $S_{3}^{(\eta)} = |\psi_R|^{2} - |\psi_L|^{2}$. These HOPS-SPs are used to construct the HOPS of index $\eta$, as illustrated in Fig. \ref{q022}. For a fully polarized singular HOPS beam, the relation $(S_{0}^{(\eta)})^2 = (S_{1}^{(\eta)})^2 + (S_{2}^{(\eta)})^2 + (S_{3}^{(\eta)})^2$ holds true. In terms of these SPs, the coordinates $2\gamma^{(\eta)}$ and $2\chi^{(\eta)}$ are given by $2\gamma^{(\eta)} = \tan^{-1}(S_{2}^{(\eta)} / S_{1}^{(\eta)})$ and $2\chi^{(\eta)} = \sin^{-1}(S_{3}^{(\eta)} / S_{0}^{(\eta)})$, respectively \cite{bansal2023stokes}.
For $\eta=0$, these HOPS-SPs reduce to the standard SPs associated with the plane wave.
\section{The inhomgeneous \textit{q}-plate}
\label{section_03}
The well-known QWP and HWP are the SU(2) birefringent medium and has a uniform retardance $\delta$ and fast axis orientation $\alpha$ in its transverse plane and hence termed as the homogeneous waveplate. If either one or both of these two defining parameters of the waveplate are spatially varying then the waveplate is known as an inhomogeneous waveplate. $q$-plate is an example of an inhomogeneous waveplate where the fast axis orientation varies azimuthally in an orderly fashion in the plane of the plate. Mathematically, the fast axis orientation is expressed as $\alpha(\phi)=q\phi+\alpha_{0}$ where $ q$ is the topological charge and $\alpha_{0}$ is the offset angle. The topological charge $q$ of a $q$-plate describes how many times the fast axis orientation rotates in a complete $2\pi$ rotation. As $\phi$ goes from $0$ to $2\pi$, the fast axis rotates by $q\times2\pi$. The SU(2) Jones matrix of the $q$-plate of retardance $\delta$ and fast axis $\alpha(\phi)$ is expressed as

\begin{equation}
M(\delta, \alpha(\phi))= 
\begin{bmatrix}
        \cos \frac{\delta}{2} + i\sin\frac{\delta}{2} \cos 2 \alpha(\phi) & i\sin\frac{\delta}{2} \sin 2 \alpha(\phi) \\[10pt]
        i\sin\frac{\delta}{2} \sin 2 \alpha(\phi) & \cos \frac{\delta}{2} - i\sin\frac{\delta}{2} \cos 2 \alpha(\phi)
\end{bmatrix}\in \text{SU(2)}.
\label{matrix01}
\end{equation}
This matrix is symmetric, satisfying the condition $M(\rho) = (M(\rho))^{T}$. Furthermore, if we set $q = 0$ in the above matrix, it reduces to the Jones matrix representing a homogeneous waveplate. The Jones matrices corresponding to the $q^{Q}$-plate and $q^{H}$-plate satisfy the eighth and fourth roots of the identity element, respectively, expressed mathematically as $(M(\pi/2), \alpha(\phi))^{8} = \mathbb{I}$ and $(M(\pi), \alpha(\phi))^{4} = \mathbb{I}$. Furthermore, the diagonal elements are complex conjugates, fulfilling
\begin{figure}[H]
\centering
\includegraphics[width=0.74\linewidth]{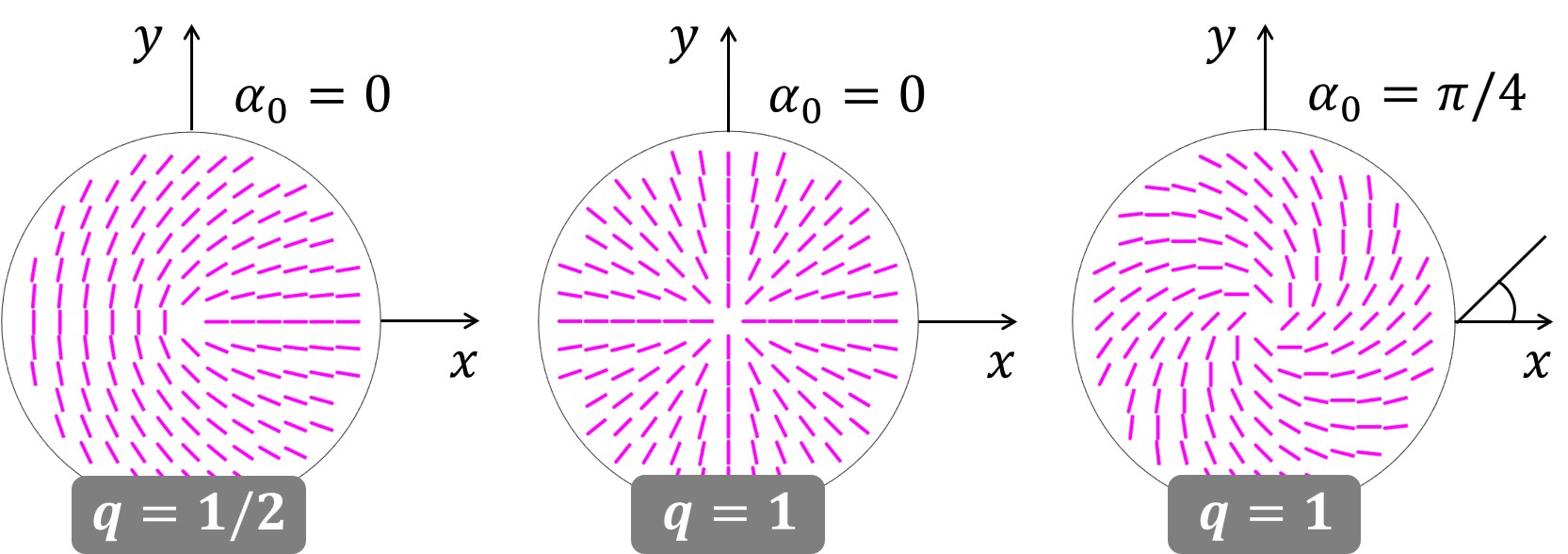}
\caption{(Color online). \textit{Geometry of the $q$-plate structure for three distinct configurations corresponding to $q=1/2$ with $\alpha_{0}=0$ (first geometry), $q=1$ with $\alpha_{0}=0$ (second geometry), and $q=1$ with $\alpha_{0}=\pi/4$ (third geometry).}}
\label{q011}
\end{figure}
\noindent
the relation  $M_{11}(\delta) = (M_{22}(\delta))^{*}$, while the off-diagonal elements are identical, i.e, $M_{12}(\delta) = M_{21}(\delta)$ and are purely real. This Jones matrix is also an SU(2) matrix $(\texttt{det}M(\delta)=1$ and $M(\delta)^{\dagger}M(\delta)=\mathbb{I})$, enabling norm-preserving transformations between two polarized optical beams. The $ q $-plate with topological charge $ q = 1 $ and different offset angles, $ \alpha_{0} = 0 $ and $ \alpha_{0} = \pi/4 $, is shown in Fig. \ref{q011}.
Further, if the Jones matrix of a waveplate is given then we can extract the information about the retardance $\delta$ and the spatially varying fast axis orientation $\alpha(\phi)$ by using expressions 
\begin{equation}
\delta = 2\cos^{-1}\Bigg[\frac{1}{2}\texttt{Tr}[M(\delta)]\Bigg], \quad
\alpha(\phi) = \frac{1}{2}\tan^{-1}\Bigg[\frac{\texttt{Im}[M_{12}(\delta)]}{\texttt{Im}[M_{11}(\delta)]}\Bigg].
\label{rhoalpha}
\end{equation}
\section{Topologically matched condition}
\label{section_04}
It is a well-known and established result that the SU(2) action of a homogeneous waveplate on uniformly polarized light can be viewed as an SO(3) rotation on the standard PS \cite{kumar2011polarization}. This connection stems from the fact that both the standard PS beam and the homogeneous waveplate share the same topology: in the beam, the polarization is uniform across the cross section, while in the waveplate, the fast-axis orientation is uniform across its transverse plane. This homogeneity represents the topologically matched condition between the PS and the waveplate. Now, we have the HOPS and the $q$-plate, both sharing the $\phi$-dependent topology of different orders and topological charges, respectively. In this section, we will determine the topological condition that enables the movement of the HOPS beam along the same HOPS using the $q$-plate.\\
\indent
When the input HOPS beam, as given in Eq. (\ref{psi}), passes through the $q$-plate, the resulting output HOPS beam is expressed as
\begin{equation}
\begin{aligned}
|\psi^{'}_{\ell}\rangle &= M(\delta, \alpha(\phi))|\psi_{\ell}\rangle \\
                          &= \psi_{1} |R\rangle + \psi_{2}|L\rangle,
\end{aligned}
\end{equation}
where, $\psi_{1}$ and $\psi_{2}$ are expressed as
\begin{equation}
    \psi_{1} = \frac{1}{\sqrt{2}}\left[i(\cos{\chi}-\sin{\chi})\sin{\frac{\delta}{2}}e^{-i\left(2\alpha_{0}-\gamma\right)}\right]e^{-i(2q-\ell)\phi}  + \frac{1}{\sqrt{2}}\left[(\cos{\chi}+\sin{\chi})\cos{\frac{\delta}{2}}e^{-i\gamma} \right]e^{-i\ell\phi},
    \label{psi_1}
\end{equation}
\begin{equation}
    \psi_{2} = \frac{1}{\sqrt{2}}\left[i(\cos{\chi}+\sin{\chi})\sin{\frac{\delta}{2}}e^{i\left(2\alpha_{0}-\gamma\right)}\right]e^{i(2q-\ell)\phi}  + \frac{1}{\sqrt{2}}\left[(\cos{\chi}-\sin{\chi})\cos{\frac{\delta}{2}}e^{i\gamma} \right]e^{i\ell\phi}.
    \label{psi_2}
\end{equation}
To ensure that the output beam $|\psi^{'}_{\ell}\rangle$ remains on the same sphere, it must respected the following form:
\begin{equation}
|\psi^{'}_{\ell}\rangle = \psi^{'}_{R}|R_{\ell}\rangle + \psi^{'}_{L}|L_{\ell}\rangle.
\end{equation}
To satisfy the above equation, it follows from Eqs. (\ref{psi_1}) and (\ref{psi_2}) that $2q - \ell = \ell$, which implies that $q = \ell = \eta$. Therefore, this condition represents the topological requirement for the output to remain on the same HOPS. If this condition is not satisfied, the output will fall on a different order of the HOPS or be located on an undefined sphere, thus failing to adhere to the intended topological structure. With this condition, the complex amplitudes $\psi^{'}_{R}$ and $\psi^{'}_{L}$, corresponding to the output can be written as
\begin{equation}
    \psi^{'}_{R} = \frac{1}{\sqrt{2}}\left[i(\cos{\chi}-\sin{\chi})\sin{\frac{\delta}{2}}e^{-i\left(2\alpha_{0}-\gamma\right)} +(\cos{\chi}+\sin{\chi})\cos{\frac{\delta}{2}}e^{-i\gamma}\right],
\end{equation}
\begin{equation}
    \psi^{'}_{L} = \frac{1}{\sqrt{2}}\left[i(\cos{\chi}+\sin{\chi})\sin{\frac{\delta}{2}}e^{i\left(2\alpha_{0}-\gamma\right)} +(\cos{\chi}-\sin{\chi})\cos{\frac{\delta}{2}}e^{i\gamma}\right].
\end{equation}

\section{Global and local rotation}
\label{section_05}
\begin{figure}{htbp}
\centering
\includegraphics[width=0.55\linewidth]{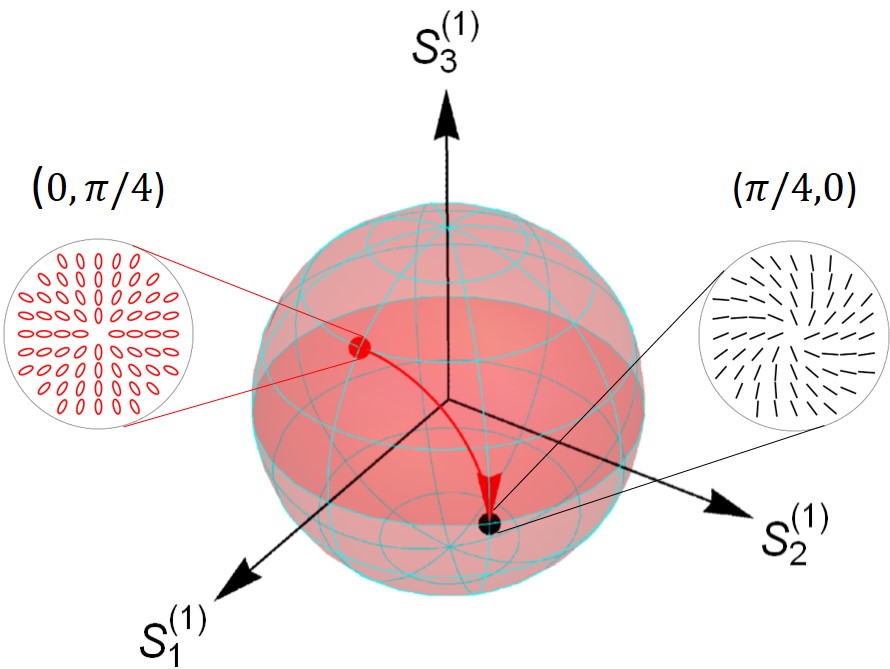}
\caption{(Color online). \textit{Transformation of an input HOPS ($\eta = 1$) beam with coordinates $(2\gamma^{(1)}, 2\chi^{(1)}) = (0, \pi/4)$ into an output HOPS ($\eta = 1$) beam with coordinates $(2\gamma^{(1)}, 2\chi^{(1)}) = (\pi/4, 0)$, induced by a $q^{Q}$-plate with offset angle $\alpha_{0} = 0$. This transformation corresponds to an SO(3) rotation on the HOPS.}}
\label{q033}
\end{figure}

\begin{figure}[t]
\centering
\includegraphics[width=1.0\linewidth]{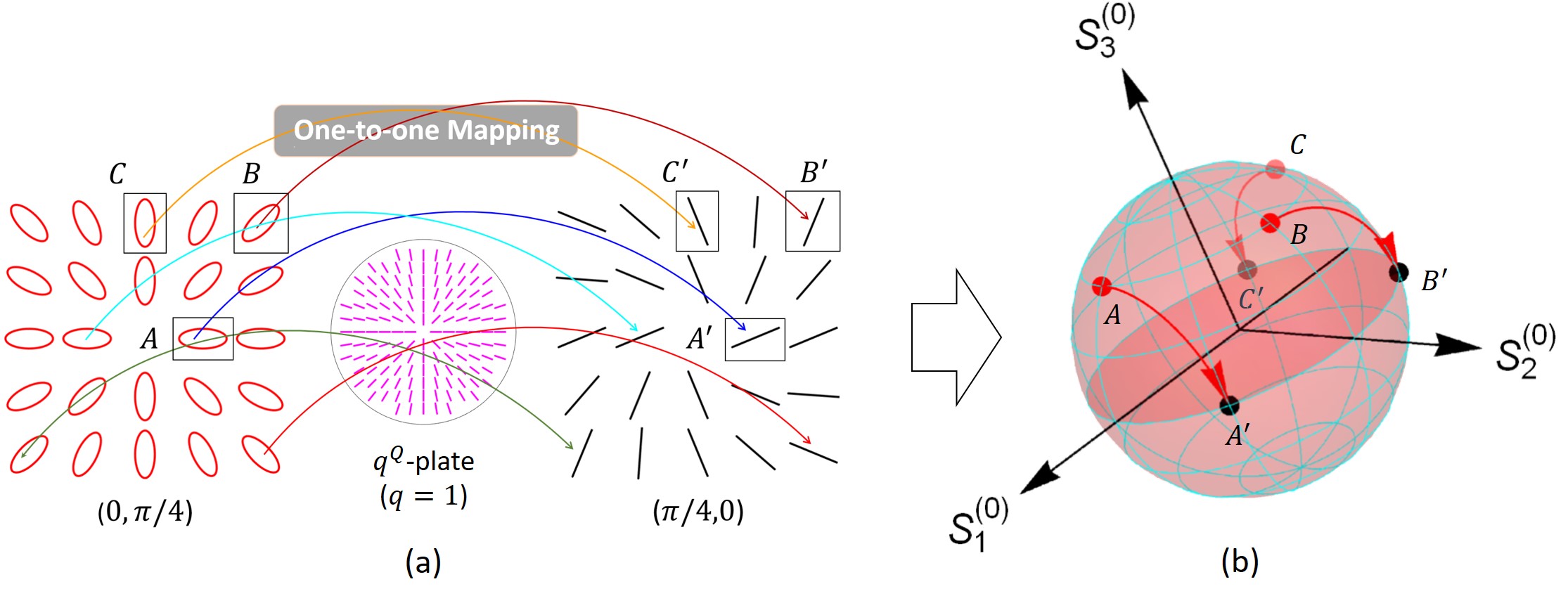}
\caption{(Color online). \textit{(a) A magnified view of the polarization evolution on the HOPS as illustrated in Fig. \ref{q033}. Each SOP of the input beam maps uniquely to an SOP of the output beam through a one-to-one correspondence, mediated by a $q^Q$-plate with an offset angle $\alpha_0 = 0$. (b) Illustration of this one-to-one mapping as an SO(3) rotation on the standard PS. The rotation occurs by an angle of $\pi/2$, with the rotation axis determined by the $\phi$-dependent fast axis orientation of the $q^Q$-plate.}}
\label{q044}
\end{figure}

When the topological condition is met, the input and output beams will remain on the same sphere, and this transformation can be interpreted as a rotation. Leveraging this, we will explore the rotation on the HOPS and demonstrate how it is connected to the rotation on the standard PS. Consider the geometry depicted in Fig. \ref{q033}, where an input beam characterized by coordinates $(2\gamma^{(1)}, 2\chi^{(1)}) = (0, \pi/4)$ passes through a $q^{Q}$-plate with an offset angle $\alpha_{0}=0$. Upon exiting the $q^{Q}$-plate, the beam coordinates transform to $(2\gamma^{(1)}, 2\chi^{(1)}) = (\pi/4, 0)$. This transformation corresponds to an SO(3) rotation about the $S_{1}^{(1)}$-axis (the rotation axis determined by $\alpha_{0}$) by an angle of $\pi/2$, which equals the retardance introduced by the $q^{Q}$-plate.  Referring to the geometry illustrated in Fig. \ref{q044}, we examine an enlarged view of the input and output beams shown in Fig. \ref{q033}. The HOPS beam consists of numerous individual SOPs, each with constant ellipticity but varying azimuthal. We highlight three specific SOPs from the input beam, labeled as $A$, $B$ and $C$. As these SOPs pass through the $q^{Q}$-plate, they individually transform into the output SOPs, denoted as $A'$, $B'$ and $C'$, respectively, thereby demonstrating a direct one-to-one correspondence between the input and output SOPs.

In Fig. \ref{q044}(a), the SOP at point $A$ is right-elliptically polarized light of zero azimuth with coordinates $(2\gamma^{(0)}, 2\chi^{(0)}) = (0, \pi/4)$ on the standard PS. When this SOP passes through the $q^{Q}$-plate, it then encounters the plate where its fast axis locally aligned horizontally, resulting in an output at $A'$ that is linearly polarized with coordinates $(2\gamma^{(0)}, 2\chi^{(0)}) = (\pi/4, 0)$. This transformation corresponds to an SO(3) rotation about the $S_{1}^{(0)}$-axis (clockwise when viewed from $S_{1}^{(0)}$ side) by an angle equal to the retardance of the $q^Q$-plate as depicted in Fig. \ref{q044}(b). Similarly, the SOP at points $B$ and $C$ encounters the $q^{Q}$-plate, where the fast axes are locally oriented at $\pi/4$ and $\pi/2$, respectively. As a result, the output polarization states appear at $B'$ and $C'$. These SU(2) transformations correspond to rotations on the PS: a clockwise rotation about the $S_2^{(0)}$-axis (when viewed from the $S_2^{(0)}$ side) and a clockwise rotation about the $S_1^{(0)}$-axis (when viewed from the $-S_1^{(0)}$ side), respectively, as depicted in Fig.~\ref{q044}. The SU(2) transformation through the waveplate space as an SO(3) rotation on the standard PS space is discussed in detail in \cite{kumar2011polarization}. Similarly, each input SOP of the HOPS beam undergoes a one-to-one mapping after passing through the $q^{Q}$-plate, owing to the fact that the beam and the plate share the same topology ($q=\eta$). Therefore, a single \textit{global} SO(3) rotation on the HOPS comprises many \textit{local} SO(3) rotations on the PS. This feature is not exclusive to the $q^{Q}$-plate but also holds true for a general $q$-plate with any retardance value between $0$ and $2\pi$. Furthermore, this \textit{global-local} feature also applies to the HOPS beam of any order $\eta$.

\noindent
\textbf{Role of $q\phi$ and $\alpha_{0}$ term:} As explained, a single global rotation on the HOPS can be viewed as a collection of multiple local rotations on the standard PS. Hence, a global rotation on the HOPS requires one rotation axis, whereas multiple local rotations necessitate multiple corresponding rotation axes. In both cases, the rotation axes are determined by the $q$-plate. The fast axis orientation of the $q$-plate is given by $\alpha(\phi) = q\phi + \alpha_0$, comprising two components: a $\phi$-dependent term $q\phi$ and a $\phi$-independent term $\alpha_0$. The constant term $\alpha_0$ specifies the rotation axis for the single global rotation on the HOPS, with the rotation axis forming an angle of $2\alpha_0$ with the $S_1^{(\eta)}$-axis in the equatorial plane of the HOPS \cite{yao2023quantitative}. The $\phi$-dependent term, $q\phi$, dictates the orientation of the multiple local rotation axes, corresponding to the multiple local rotations on the standard PS. It is important to note that, whether on the HOPS or PS, the rotation axis lies within the equatorial plane of the sphere.
\section{SU(2) polarization evolution on higher-order Poincar\'{e} sphere}
\begin{figure*}[t]
\centering
\includegraphics[width=1.0\linewidth]{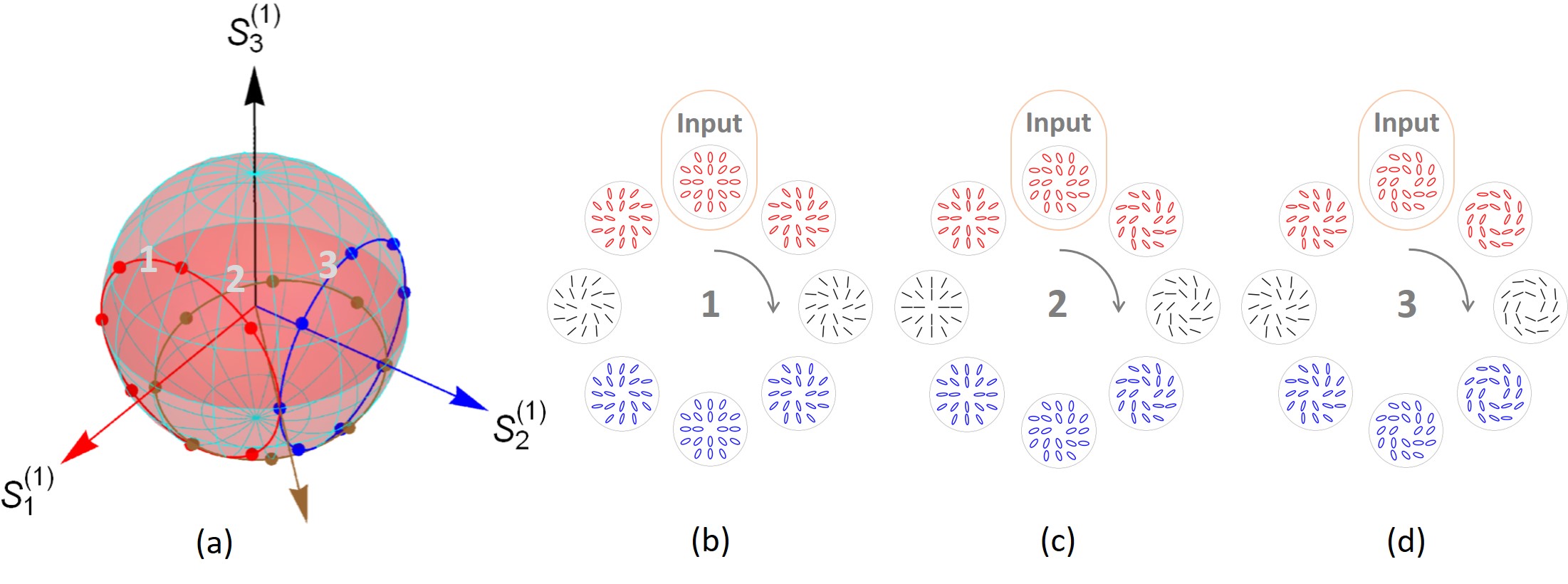}
\caption{(Color online). \textit{(a) Three circular trajectories labeled as $1$, $2$ and $3$ are shown on the HOPS for $\eta = 1$, with points placed at angular intervals of $\pi/4$ at each trajectory. For each circle, the polarization evolution is demonstrated by applying a $q$-plate to an input beam while varying the retardance from $0$ to $2\pi$ in steps of $\pi/4$. The $q$-plates corresponding to circles $1$, $2$ and $3$ have offset angles $\alpha_0 = 0$, $\pi/8$, and $\pi/4$, respectively.
Figs. (b), (c) and (d) show the resulting polarization evolution for circles $1$, $2$ and $3$, respectively. The rotation axis makes an angle of $2\alpha_0$ with the $S_1^{(1)}$-axis, while the rotation angle is governed by the retardance. It is important to note that there is no correlation between the red and blue colors of the circular trajectory and the red and blue colors of the polarization states.}}
\label{q055}
\end{figure*}

In the previous section, we demonstrated the connection between the rotation on HOPS and the rotation on PS, providing a comprehensive description of the rotation dynamics on HOPS. In this section, we will discuss the rotation on HOPS using a general $q$-plate. The general $q$-plate extends the conventional $q$-plate by incorporating a continuously tunable retardance and offset angle, which vary from $0$ to $2\pi$ and $0$ to $\pi/2$, respectively.\\
\indent
Now, consider the figure illustrated in Fig. \ref{q055}. Fig. \ref{q055}(a) illustrates the geometry of the HOPS for $\eta = 1$ with three circular trajectories with points placed at angular intervals of $\pi/4$. Fig. \ref{q055}(b) presents a two-dimensional projection of circle $1$ along with the corresponding polarization distribution. For this trajectory, the input beam is chosen with coordinates $(2\gamma^{(1)}, 2\chi^{(1)}) = (0, \pi/4)$, and the $q$-plate has an offset angle of $\alpha_0 = 0$. As the input beam passes through the $q$-plate with increasing retardance from $0$ to $2\pi$ in steps of $\pi/4$, the output beam traces the path of circle 1. In this case, the rotation is clockwise when viewed along the $S_{1}^{(1)}$-axis. In Fig.~\ref{q055}(b), the input beam has coordinates $(2\gamma^{(1)}, 2\chi^{(1)}) = (\pi/4, \pi/4)$ and passes through a $q$-plate with an offset angle of $\alpha_0 = \pi/8$. Again, when the retardance varies from $0$ to $2\pi$ in steps of $\pi/4$, the output beam follows the trajectory of circle 2, rotating clockwise about an axis inclined at $2\alpha_0 = \pi/4$. Fig. \ref{q055}(c) illustrates a similar evolution for a different input beam with coordinates $(\pi/2, \pi/4)$ and a $q$-plate having offset angle of $\alpha_0 = \pi/4$. In this case, the output beam rotates clockwise around the $S_{2}^{(1)}$-axis and traces the path of circle 3, as viewed along the $S_{2}^{(1)}$-axis. Here, the rotation axis is the $S_{2}^{(1)}$-axis, which forms an angle of $\pi/2$ with the $S^{(1)}_{1}$, as in this case, $2\alpha_{0} = \pi/2$. Hence, this simulation experiment demonstrates that, under the topological condition, the $q$-plate plays a crucial role in governing the polarization evolution on the HOPS. This feature also holds true for all order of HOPS. 
\label{section_06}
\section{Conclusion}
\label{section_07}
In this paper, we have discussed three key aspects in the context of HOPS and the $q$-plate: the topological condition, global-local rotation, and the SU(2) polarization evolution on the $\eta$-order sphere. First, we examined the topologically matched condition between the order of the HOPS and the topological charge of the $q$-plate. Under this condition, the input and output states remain on the same world when the transformation occurs via a general $q$-plate. Leveraging this, we demonstrated that a single global SO(3) rotation on the HOPS, induced by an SU(2) action of the $q$-plate, is a collection of multiple local SO(3) rotations on the standard PS. This \textit{global-local} connection applies to the HOPS beam of any order $\eta$. Analogous to the standard PS, where the rotation axis in the equatorial plane is determined by the fast-axis orientation of the waveplate, in the case of HOPS, the offset angle $\alpha_0$ of the $q$-plate determines the rotation axis in the equatorial plane of the HOPS. Furthermore, the $q\phi$ term governs the orientation of multiple local rotation axes across the standard PS. Notably, our findings show that with a $q$-plate capable of tuning its retardance from $0$ to $2\pi$ and its offset angle from $0$ to $\pi/2$, complete SU(2) polarization evolution is achievable on the HOPS. This level of tunability can be realized through modern technologies such as metamaterials, metasurfaces, and liquid crystal-based devices, enabling precise control over polarization evolution. This level of control enables the design of reconfigurable optics for manipulating structured light.\\
\\
\\
\noindent
\textbf{Funding:} Science and Engineering Research Board (SERB) India (CRG/2022/001267).\\
\\
\noindent
\textbf{Disclosures:} The authors declare that there are no conflicts of interest.\\
\\
\noindent
\textbf{Acknowledgement:} MU acknowledges the support of the Fellowship from IIT Delhi. MU also expresses sincere gratitude to the members of the Singular Optics Lab at IIT Delhi for their invaluable support and encouragement throughout the research.\\
\\

\newpage
\bibliographystyle{elsarticle-num}
\bibliography{References}

\end{document}